\documentclass[sigconf, nonacm]{acmart} % Use only one \documentclass declaration
\usepackage{array} % For custom column formatting
\usepackage{graphicx} % For \resizebox if used
\usepackage{lscape} % Optional, for landscape orientation if needed
\usepackage{afterpage} % Include this in the preamble

\begin{document}
\title{A Simple and Fast Way to Handle Semantic Errors in Transactions}

%
% The "author" command and its associated commands are used to define the authors and their affiliations.
\author{Jinghan Zeng}
\affiliation{%
  \institution{The University of Chicago}
}
\email{zengjinghan@uchicago.edu}

\author{Eugene Wu}
\affiliation{%
  \institution{Columbia University}
}
\email{ewu@cs.columbia.edu}

\author{Sanjay Krishnan}
\affiliation{%
  \institution{The University of Chicago}
}
\email{skr@uchicago.edu}

\begin{abstract}
Many computer systems are now being redesigned to incorporate LLM-powered agents, enabling natural language input and more flexible operations. This paper focuses on handling database transactions created by large language models (LLMs). Transactions generated by LLMs may include semantic errors, requiring systems to treat them as long-lived. This allows for human review and, if the transaction is incorrect, removal from the database history. Any removal action must ensure the database's consistency (the "C" in ACID principles) is maintained throughout the process.

We propose a novel middleware framework based on Invariant Satisfaction (I-Confluence), which ensures consistency by identifying and coordinating dependencies between long-lived transactions and new transactions. This middleware buffers suspicious or compensating transactions to manage coordination states. Using the TPC-C benchmark, we evaluate how transaction generation frequency, user reviews, and invariant completeness impact system performance. For system researchers, this study establishes an interactive paradigm between LLMs and database systems, providing an "undoing" mechanism for handling incorrect operations while guaranteeing database consistency. For system engineers, this paper offers a middleware design that integrates removable LLM-generated transactions into existing systems with minimal modifications.

\end{abstract}

\maketitle

\section{Introduction}

For decades, system design has been centered around humans as the primary users and operators. In database management systems (DBMS), transactions are usually initiated and committed by human users. While syntactic errors can be easily detected and corrected through standard error handling, users may also introduce semantic errors or need additional time to decide whether to commit a transaction. To manage these challenges, researchers and engineers have developed "undo" mechanisms, ensuring both data integrity and system recoverability.

The ARIES protocol (Algorithm for Recovery and Isolation Exploiting Semantics) has long been the gold standard for recovery, combining write-ahead logging, undo, and redo operations to efficiently manage fine-grained rollbacks \cite{mohan1992aries}. However, managing semantic errors requires transactions to remain long-lived for user review, making ARIES potentially too storage-intensive for such use cases. Beyond ARIES, alternative approaches such as sagas have been developed for long-running transactions in distributed systems, enabling partial rollbacks through compensating transactions while maintaining system consistency \cite{garcia1987sagas}. ACTA is derivative work for sagas, and it extends sagas to include extended transactions while still providing compensating transactions and maintaining consistency\cite{chrysanthis1992acta}. Similarly, advancements in multi-version concurrency control (MVCC) allow systems to maintain historical states, enabling rollbacks to previous valid database states \cite{bernstein1987concurrency}. Escrow methods, often utilized in high-concurrency environments, only allowed conditional updates and guarantees to facilitate safe rollbacks without locking resources \cite{o1986escrow}. These mechanisms either allow "undo" or give users a chance to store long-lived transactions and review the transactions to fix semantic errors.

Large Language Models (LLMs) are increasingly being seen as active operators within systems, either by translating human commands for system interaction or functioning autonomously to maintain the system based on predefined prompts. LLMs have demonstrated their ability to understand system structures and effectively query data \cite{zhang2024llmssubstitutesqlcomparing, 10.14778/3685800.3685916, Fernandez2023HowLL, chen2024seeddomainspecificdatacuration}. While these actions typically do not modify the original database state and instead provide more accessible methods for data extraction, more complex interactions involve "write" actions that alter the database state \cite{patil2024goex, subramaniam2024intentbasedaccesscontrolusing, patil2023gorilla, wornow2024automatingenterprisefoundationmodels, zhang2024llmssubstitutesqlcomparing}. These write actions can introduce errors, which are often manageable through validation, as well as semantic errors, which may require user review for correction \cite{zhang2024llmssubstitutesqlcomparing, xu2024hallucinationinevitableinnatelimitation, patil2023gorilla}.

An intuitive solution is to improve the prediction accuracy and reduce semantic errors of LLMs in adhering to user instructions and expectations. However, due to inherent limitations in LLMs, achieving perfect performance is unrealistic. Thus, we must treat incorrect transactions as errors and remove them when necessary. Existing approaches like ARIES, sagas, escrow, and MVCC provide partial solutions but fall short of addressing the challenges posed by long-lived transactions, especially those generated by LLMs. ARIES, for instance, does not support long-lived transactions. Sagas undo transactions by predefined independent compensating transactions, which cannot handle inter-dependent transactions that violate system constraints. Sagas also requires system implementation case by case. ACTA briefly explains how to handle transactions that cannot be undone. However, it does not discuss inter-dependent transactions, which may violate database constraints based on the other transaction's state, not just an independent transaction itself. Escrow focuses on counter-type data, limiting its applicability. MVCC does not support selective removal of specific transactions. Approaches like rewriting history rely on commutativity checks to remove incorrect transactions and transactions affected by incorrect transactions, but commutativity imposes overly strict constraints \cite{liu2000rewriting}. A more practical approach would relax this standard by enforcing database constraints rather than insisting on identical database states. Thus, we adapted and extended Invariant Confluence to check whether LLM-generated transactions require review and may need coordination with newer transactions. \cite{bailis2014coordination}. Coordinations stall certain newer transactions to guarantee that the undoing of long-lived transactions always leads to a consistent database state. We also reviewed additional papers on long-lived transactions; see the details in the discussion section.
\begin{table}[ht]
    \centering
    \caption{Comparison of Undo Strategies}
    \resizebox{0.5\textwidth}{!}{  % Resize the table to half of the page width
        \begin{tabular}{|c|c|c|c|c|c|}
            \hline
            \textbf{Feature} & \textbf{ARIES} & \textbf{Sagas} & \textbf{ACTA} & \textbf{Escrow} & \textbf{Proposed Method} \\ \hline
            Long-Lived Undo &  & Yes & Yes & Yes & Yes \\ \hline
            Unified Implementation & Yes  &  &  & Yes & Yes \\ \hline
            Dynamic Recoverability &  &  &  & Yes & Yes \\ \hline
            Diverse Constraints & Yes  & Yes & Yes &  & Yes \\ \hline
        \end{tabular}
    }
    \label{tab:comparison_undo_strategies}
\end{table}

In this paper, \textbf{Section 2} outlines the system settings and requirements. \textbf{Sections 3 and 4} summarize the processes for identifying transactions that may require coordination for the consistency of long-lived transactions waiting for reviewing. \textbf{Sections 5 and 6} explain the roles of transaction managers and middleware in supporting coordination mechanisms. \textbf{Section 7} explores the interplay between system availability, human review, LLM-generated transactions, and the completeness of query and invariant information. \textbf{Section 8} compares and discusses various strategies, highlighting their strengths and limitations.

At a high level, developers using this framework must first manually analyze the dependencies between each pair of SQL query templates used to interact with their database models. The dependency check details for each pair can be referenced in Table 4. After identifying the dependencies, developers need to register them in the dependency-checking function within the middleware (Section 6). Once a transaction request is received from the user, the middleware passes the dependency-checking function along with the new request to the transaction manager (Section 5). The transaction manager then asynchronously coordinates the transactions. Users can either check the status of their submitted transactions or review LLM-generated transactions (Section 6).

For researchers working on LLM-data system integration, they can focus on each component separately for optimization.

\section{System Setting}
\begin{figure}
    \centering
    \fbox{\includegraphics[width=\linewidth]{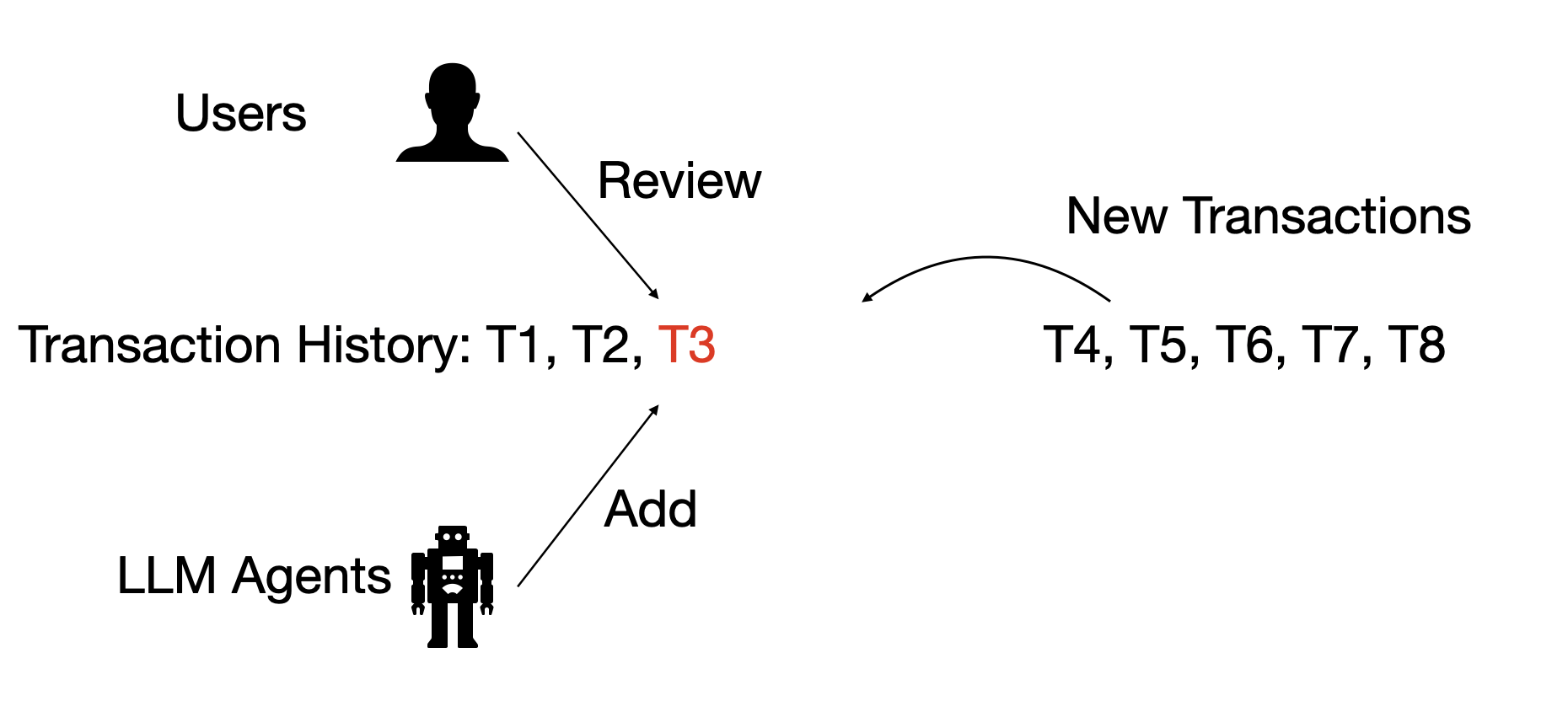}}
    \caption{System Setting}
    \label{fig:sys_setting}
\end{figure}

We consider a system that operates in an environment that accepts transactions from human users and LLMs (Large Language Models). Occasionally, transactions may later be identified as incorrect by users, such as when users decide they no longer want to proceed with a transaction or when a semantic error in LLM-generated transactions leads to an imperfect execution of users' instructions. For example, in Figure \ref{fig:sys_setting}, after transaction 2 (T2), an account balance is 50 USD, and transaction 3 (T3) increments 10 USD while all new transactions T4 - T8 have a net effect that decreases by 60 USD. One database constraint requires that the account balance must be positive. T3 is an LLM-generated transaction; the user reviewed it and decided to remove it. In this case, T3 cannot be removed; otherwise, it will lead to a violation of the database constraint. This setting raises two key questions:

\begin{enumerate}
    \item How can we isolate or remove these questionable transactions from LLMs, especially when they may no longer be needed, from database history?
    \item     In certain situations, removing a suspicious transaction might not be feasible while maintaining the consistency of the data system due to subsequent new transactions' modification to the database.
\end{enumerate}

\textit{Solution for (1): In this paper, we mainly suggest two approaches for suspicious transaction removal/separation, and we discussed the problem of other approaches: }\\

\textbf{Buffering Suspicious Transactions:} Rather than committing suspicious transactions directly to the database, the web developer can store these transactions in a buffer. Once an administrative user reviews and either accepts or removes the transaction, the transaction manager within the system can then commit or discard the transaction accordingly.

\textbf{Buffering Compensating Transactions:} A compensating transaction is a specialized transaction used to "undo" the effects of a previously committed transaction by applying an opposite action to restore the system to a consistent state \cite{korth1990formal}. Unlike a rollback, which cancels a transaction before it is committed, a compensating transaction is applied after commitment to address cases where reversal is necessary without altering the original transaction history. For instance, in a banking system, if 100 USD is mistakenly transferred from Account A to Account B, the compensating transaction would transfer 100 USD back to Account A to correct the error. Compensating transactions is essential in complex systems where direct rollbacks may be impractical because the transaction has already been committed.

In our context, the system commits the transaction upon receipt and buffers its corresponding compensating transaction. If an administrative user reviews and decides to remove the transaction, the transaction manager within the system can then commit a compensating transaction to negate the effect of the suspicious transaction on the database. If the transaction is approved, no further action is needed for the database.\\

To ensure the database maintains consistency while also allowing for the removal or separation of suspicious transactions, we consider several approaches: 

\textbf{1) Full Database Locking}: The simplest approach is locking the entire database whenever a suspicious transaction request is received. The system continues running but restricts other transactions until users review the suspicious transaction. This enforces strict serializability and ensures ACID properties, particularly consistency. \textbf{2) Sandbox Simulation}: Another option is to create a sandbox to simulate transactions. As long as a compensating transaction can remove the effects of suspicious transactions while maintaining consistency, even after subsequent transactions, the suspicious transaction is safe to undo. \textbf{3) Granular Locking}: Locking specific database parts (e.g., rows or tables) can also support removability. However, determining the appropriate granularity to maintain removability without compromising consistency is challenging. \textbf{4) Naive Buffering of Suspicious/Compensating Transactions}: Suspicious transactions can be temporarily buffered without immediate execution. However, without coordination, a buffered transaction that is initially valid may become uncommittable due to subsequent modifications. Each of these approaches has limitations, which we will further discuss in Section 8.

5) \textbf{Buffering with Logical Dependency Checks}: A logical dependency check can identify new transactions that would compromise the validity of undoing a buffered transaction without committing the buffered transaction. By holding these identified transactions, the system can ensure that all buffered transactions remain valid. Specifically, if the system has knowledge of database invariants and transaction conditions, an \textbackslash{}textit\{Invariant Satisfaction\} (or \textbackslash{}textit\{Invariant Confluence\}) analysis can assess dependencies between buffered transactions and new transactions. Whether buffering suspicious transactions or compensating transactions, invariant analysis determines if new transactions should proceed or be held, maintaining consistency.\\

\textit{Solution for (2):  We identify new transactions that will lead to the inconsistency; we will provide inconsistency detection in section 3:}\\

\textbf{Consistency:} Consistency in this paper refers to the "C" in the ACID property of database transactions \cite{gray1981transaction,haerder1983principles}. From the perspective of the database, the developers must guarantee its consistency; specifically, they need to ensure the predefined constraints from the table creators are not violated.

\textbf{Dependency:} We define dependency as follows: when two transactions arrive in sequence, the system may be unable to accept both due to consistency constraints. In this case, the second transaction is said to depend on the first. For example, consider a database constraint requiring an account balance to remain above 0 and assume an account currently holds 50 USD. If the first transaction deducts 40 USD and is marked as suspicious, it may be buffered for possible removal. However, if a second transaction attempts to deduct 20 USD, it will be removed or held, as committing both would violate the balance constraint. Thus, the second transaction depends on the first and requires coordination, such as buffering, to maintain consistency.

\textit{Recoverability:} In the buffering compensating transaction approach, when a transaction commits to the database, it transforms the database from state 1 to state 2. If we later need to undo this transaction’s changes, we must apply a compensating transaction defined by the system designer. However, there is a risk that a compensating transaction could result in an inconsistent state. We define the ability to remove the committed Transaction 2 with consistency satisfaction as recoverability. In a system that buffers suspicious transactions, the buffered suspicious transaction can always be safely removed without affecting a consistent system state. From the perspective of buffering compensating transactions, the recoverability of a suspicious transaction is decided by its corresponding compensating transaction's dependency on a new transaction. If a committed new transaction does not depend on its compensating transaction, the suspicious transaction can be recovered.

\textbf{Availability: }Availability is impacted by dependencies between transactions. When a suspicious or compensating transaction is buffered, new transactions that depend on it may also be held. The system will delay any dependent transactions until the buffered transaction receives final approval from administrators or users. This holding process reduces transaction throughput and affects overall system availability.

\textbf{CAD theorem:} There are trade-offs between consistency, availability, and dependency. In our system setting, we cannot achieve the three of them at the same time. This is similar to the CAP theorem in distributed systems \cite{gilbert2002brewer}.
\begin{itemize}
    \item \textbf{Consistency and Availability:} To maintain both, we need to minimize the duration of dependencies by reducing the time that suspicious transactions are held.
    \item \textbf{Consistency and Dependency:} Achieving both often requires sacrificing some availability, which may involve reducing the number of accepted transactions and delaying transaction completion within a defined time frame.
    \item \textbf{Availability and Dependency:} To maintain availability and dependency, we must relax transaction consistency requirements. Specifically, fewer database or application constraints should be enforced at the system design stage.
\end{itemize}

\subsection{Middleware}
For web developers and system engineers, introducing a new framework in a modern enterprise is often discouraged due to concerns about cost and system reliability. Frameworks like LangChain address this by offering tool packages that integrate into larger systems as microservices \cite{langchain}. Similarly, this project focuses on creating middleware compatible with existing web frameworks, minimizing disruption to current workflows. By avoiding changes to the database or data schema, our solution ensures data integrity and reliability without requiring complex system migrations.

The Model-View-Controller (MVC) \cite{burbeck1992applications} is a widely used design pattern in software development, especially for web applications, which divides an application into three interconnected components: Model, View, and Controller. The \textbf{Model} manages data and business logic, representing the application’s core functionality by handling data storage, retrieval, and processing. The \textbf{View} serves as the presentation layer, displaying data from the Model to the user and relaying user commands to the Controller. The \textbf{Controller} acts as an intermediary, processing user inputs, updating the Model, and instructing the View to display the updated data.

\begin{figure}
    \centering
    \fbox{\includegraphics[width=1\linewidth]{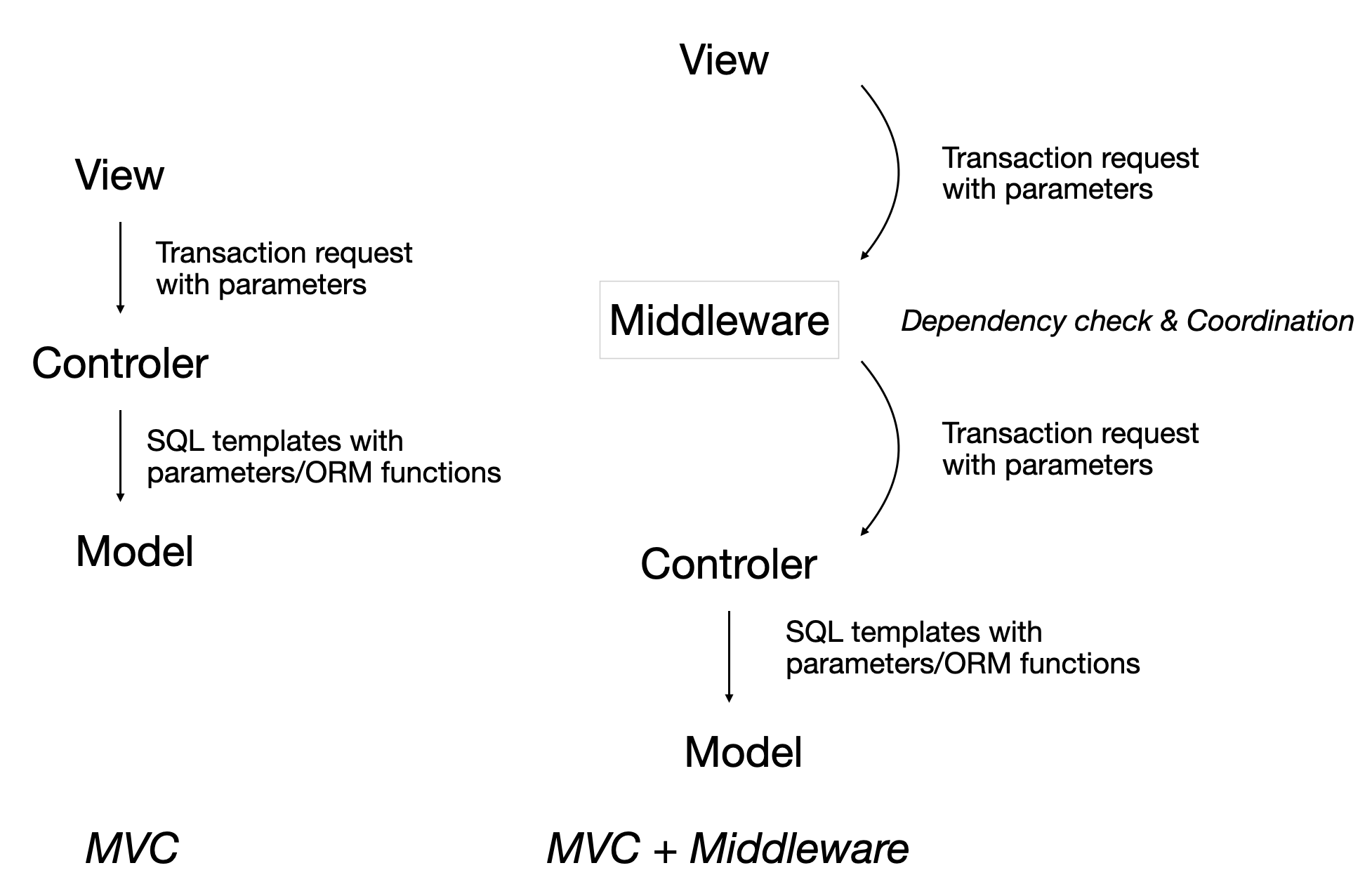}}
    \caption{MVC  with Middleware}
    \label{fig:enter-label}
    
\end{figure}

In our setting (in Figure \ref{fig:enter-label}), after the view receives the request from either users or LLM agents, it will not directly talk to the controller. Instead, the control workflow will be delegated to the middleware, which ensures that new transactions that depend on buffered transactions can be appropriately coordinated to maintain system consistency.

As we discussed before, there are two approaches to buffering: buffering suspicious transactions and buffering compensating transactions. We will use the second one as an example to explain how the middleware works.

\subsection{Dependency-Check for a Middleware}
We assumed web development based on MVC, and the middleware only received the transaction name and parameters. Usually, a transaction request triggers predefined transaction logic according to the transaction name and parameters (details in section 3). The web developer needs to pre-analyze that logic's dependency and register it in the \textit{dependency check function}. In the runtime, if the server receives one new transaction, it can always check its dependency with former transactions with the \textit{dependency check function}.

\subsection{Buffering Compensating Transactions}

\subsubsection{Request from the User}
Assume that the User or LLM interacts with the system via an HTML interface, with actions on the webpage submitted through URL requests. Rather than directly calling database systems or ORM operations, these URL-based requests are first placed into a queue, where the Transaction Manager processes them periodically. Each transaction request is assigned a unique transaction ID before entering the queue. This ID is also logged or updated in the transaction status table with a status of “submitted,” which allows for universal tracking and identification of each transaction's status.

In a separate process running parallel to the user-facing request handler, the Transaction Manager periodically retrieves requests from the queue, assigning each request for dependency check. The Transaction Manager (details in section 4) decides whether a new transaction should be buffered or committed based on its dependency on buffered transactions. It also checks if any transactions in the buffer are ready to be removed or executed.

\subsubsection{Review a Buffered Transaction}
If an admin or user identifies a suspicious buffered LLM-generated transaction, they can choose to accept or remove it. Once the approval or removal is triggered and submitted to the system via a URL request, the request is added to a decision buffer.

During each iteration, the Transaction Manager retrieves results from the decision buffer and removes the corresponding buffered transaction. It then commits these removed or “mature” buffered transactions and updates the status table accordingly before accepting new transactions.

\subsubsection{Checking the Status}
When a request for transaction status is received, the system retrieves the result or status from the status table using the transaction ID.

\subsection{Buffering Suspicious Transactions}
Same framework except for buffering suspicious transactions instead of compensating transactions.

\subsection{API Documentation}

\begin{figure}[h]
    \centering
    \includegraphics[width=0.5\textwidth]{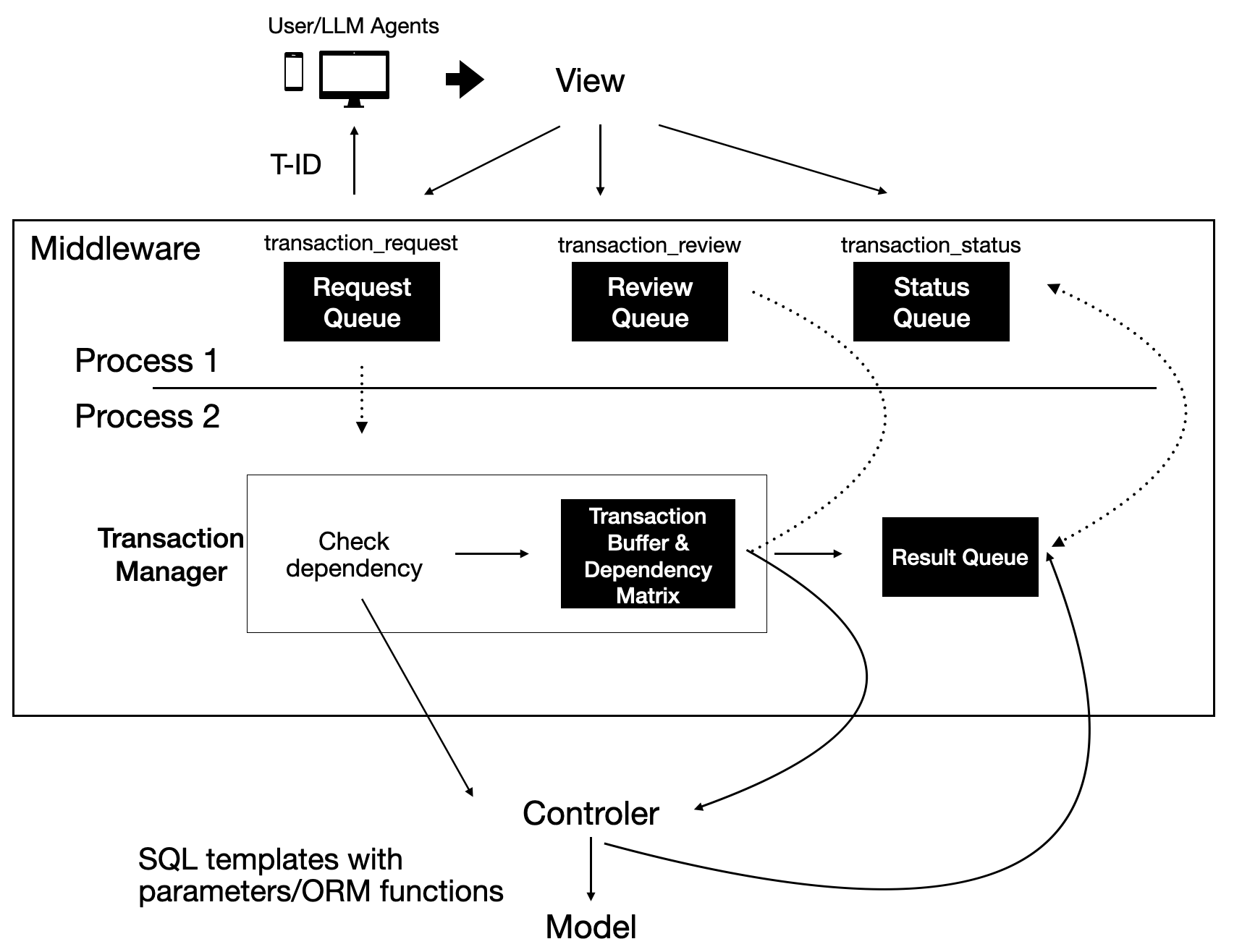}
    \caption{Web Development Middleware Framework}
    \label{fig:django}
\end{figure}

\begin{enumerate}
    \item \textbf{Endpoint:} \texttt{transaction\_request}
    
    \textbf{Description:} Initiates a transaction request and returns a unique transaction ID. \\
    \textbf{HTTP Method:} \texttt{POST}
    
    \textbf{Parameters:}
    \begin{itemize}
        \item \texttt{transaction\_name} (string): The name of the transaction.
        \item \texttt{transaction\_parameters} (object): Specific parameters required for the transaction.
    \end{itemize}
    
    \textbf{Returns:}
    \begin{itemize}
        \item \texttt{transaction\_id} (string): A unique identifier for the requested transaction.
    \end{itemize}

    \item \textbf{Endpoint:} \texttt{transaction\_review}
    
    \textbf{Description:} Accept or remove an LLM-generated and suspicious buffered transaction. \\
    \textbf{HTTP Method:} \texttt{POST}
    
    \textbf{Parameters:}
    \begin{itemize}
        \item \texttt{transaction\_id} (string): The unique identifier of the transaction to be processed.
    \end{itemize}
    
    \textbf{Returns:}
    \begin{itemize}
        \item \texttt{status} (string): The processing result, indicating whether the transaction has been accepted or removed.
    \end{itemize}

    \item \textbf{Endpoint:} \texttt{transaction\_status}
    
    \textbf{Description:} Retrieves the status of the database query result of a specific transaction. \\
    \textbf{HTTP Method:} \texttt{POST}
    
    \textbf{Parameters:}
    \begin{itemize}
        \item \texttt{transaction\_id} (string): The unique identifier of the transaction.
    \end{itemize}
    
    \textbf{Returns:}
    \begin{itemize}
        \item \texttt{transaction\_status} (string/object): The current status or database query result related to the transaction.
    \end{itemize}
\end{enumerate}

\section{Constraints in Long-Lived Transactions}

\subsection{Invariant Satisfaction }

\begin{figure}
    \centering
    \includegraphics[width=\linewidth]{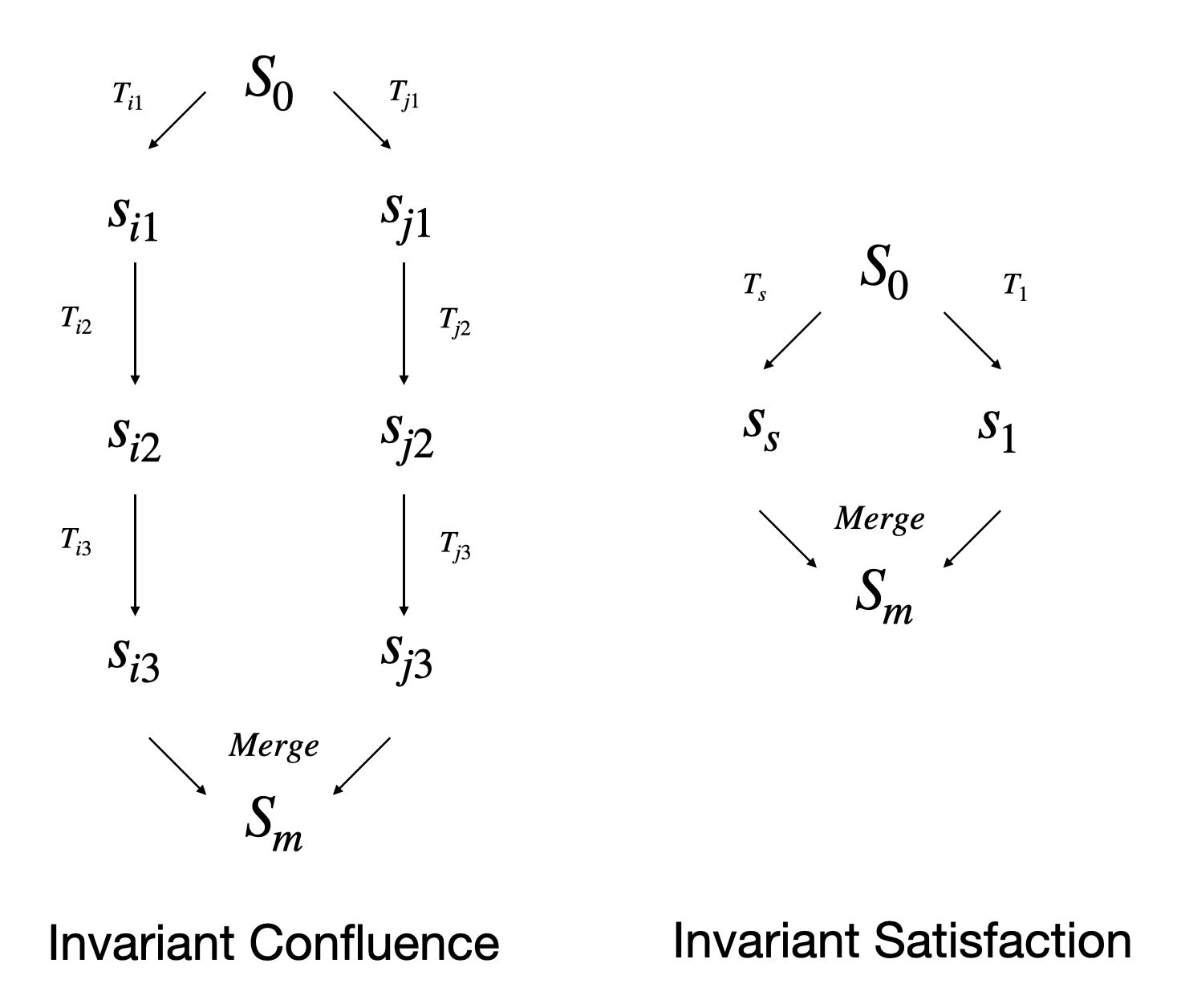}
    \caption{A comparison between Invariant Confluence (IC) and Invariant Satisfaction (IS). Here, \(S\) represents valid database states \(i\) and \(j\) denote transaction branches, \(S_s\) indicates a suspicious transaction, and \(S_1\) refers to a new transaction. The \(Merge\) function combines two states into a valid state. Coordination is required if the \(Merge\) function cannot produce a valid state.}
    \label{fig:ic-is-comparison}
\end{figure}

\textbf{Invariant confluence: } Invariant confluence \cite{bailis2014coordination, whittaker2020checking}, originally applied in distributed systems, ensures that, after merging, distributed data stores still satisfy a common invariant (i.e., database constraints). Specifically, starting from a common ancestor database state, there are two concurrent branches, each with a sequence of transactions, as shown in Figure \ref{fig:ic-is-comparison}. If each sequence results in a consistent database state (All S are consistent state) and the two branches can be merged without violating consistency, this state is called invariant confluence. 

A key theorem of invariant confluence states that a set of transactions T can execute without coordination and converge to a consistent state if and only if T is invariant confluent. Coordination here means that if two transactions are not invariant confluent, one should be held to prevent concurrent execution. In the setting of this paper and Invariant Satisfaction, coordination requires that the latter be held until the former finishes reviewing.

The invariant confluence between two transactions can be assessed by logically verifying their operations against database invariants. If two transactions are not invariant confluent and the first transaction executes earlier, we say that Transaction 2 depends on Transaction 1.

\textbf{Invariant Satisfaction: } In our setting, while a suspicious transaction remains buffered, new transactions may be committed to the database in ways that prevent the buffered transaction from committing consistently. When treating the buffered transaction and a new transaction as parallel, concurrent operations, this scenario is similar to invariant confluence. 

We propose that if the buffered transaction and the new transaction are invariant confluent, they can proceed without coordination. If they are not invariant confluent, the new transaction must be held until the buffered transaction is either accepted or removed by the user.

A key difference from traditional invariant confluence is that our setting involves only a single transaction in each "branch." We refer to this as Invariant Satisfaction (I-Satisfaction) to specify the relationship between buffered and new transactions. This also implies that if the transactions are not I-satisfied, and the buffered transaction occurs before the new transaction, then the new transaction depends on the buffered transaction.

\subsection{Invariants Analysis}
In this section, we outline a series of invariants that web developers should consider. We list and explain pairs of invariants and operations that do not meet Invariant Satisfaction requirements and, therefore, require coordination. 
\textit{(assuming 1. Transaction A is a suspicious transaction that might be removed and not exist; 2. Transaction B is a new transaction; 3. Transaction A occurs before Transaction B; 4. Transaction A and Transaction B are not committed to the database for a constraint validation check but are logically analyzed for dependency)}\\

\textbf{1 Database Constraints:} These are built-in constraints within relational database systems. This section examines cases where these constraints do not meet Invariant Satisfaction requirements, necessitating coordination \cite{chamberlin1974sequel}.

\textbf{UNIQUE}: Ensures all values in a column are unique. \textit{Non-Invariant Satisfaction Example: }Consider a database table of user emails. If Transaction A inserts a user with the email \texttt{"user@example.com"} and Transaction B also inserts a user with the same email concurrently, merging these states would violate the uniqueness constraint. Coordination is required to prevent such conflicts.

\textbf{FOREIGN KEY}: Maintains integrity by preventing actions that would destroy links between tables. \textit{Non-Invariant Satisfaction Example: }Suppose Transaction A inserts a new employee record with a foreign key reference to a department, and Transaction B deletes the same department concurrently. Merging these states would result in an invalid state with a "dangling pointer." Coordination ensures that dependent data remains consistent.

\textbf{Auto-Increment}: Automatically generates a unique identifier for new rows. \textit{Non-Invariant Satisfaction Example: }If Transaction A inserts a new record with ID 100, and Transaction B inserts another record concurrently with ID 100 as well, the auto-increment constraint is violated. Coordination ensures unique ID generation across transactions.

\textbf{CHECK Constraints}: Enforces that values in a column must meet a specific condition. \textit{Non-Invariant Satisfaction Example:} In a financial application, a column for account balance may have a CHECK constraint requiring that all balances be greater than or equal to zero. If Transaction A deducts \$400 from an account with a balance of \$400 and Transaction B deducts \$100 simultaneously, the merged state could result in a negative balance, violating the constraint. Coordination ensures that these operations do not breach the CHECK condition. Without knowing the balance, two decrement transactions are not I-satisfaction for > constraint; two increment transactions are not I-satisfaction for < constraint; all transactions are I-satisfaction for = constraint.\\

\textbf{2 Abstract Data Type Constraints:}\cite{conway2012logic,lynch1993atomic, shapiro2011comprehensive}

\textbf{Counter Constraints}: Similar to check constraints. \textit{Non-Invariant Satisfaction Example:} In an accounting system, assume the balance of an account must remain positive. If Transaction A deducts \$30 and Transaction B deducts \$40 simultaneously from an account with only \$50, the resulting balance after merging would be negative, violating the constraint. Coordination is needed to manage concurrent deductions.

\textbf{Set, List, Map Size Constraints}: Track the collection size or state accurately. \textit{Non-Invariant Satisfaction Example:} In a collaborative document editing system, if Transaction A adds a new section to a list and Transaction B adds another section simultaneously, the final list size or state might not reflect both changes correctly. Coordination ensures the consistency of collection updates.\\

New Constraints (Not Mentioned by the Coordination Avoidance Paper)\\

\textbf{Tree and Hierarchical Constraints}: Ensure relationships between parent and child nodes adhere to specific rules. \textit{Non-Invariant Satisfaction Example:} In a file system, Transaction A deletes a parent directory while Transaction B adds a new file to that directory. Merging these transactions without coordination would violate the hierarchical constraint, leading to an invalid state.

\textbf{Graph Constraints}: Enforce rules specific to graph-based data structures, such as avoiding circular references. \textit{Non-Invariant Satisfaction Example:} In a social network, Transaction A adds a relationship where User A follows User B, and Transaction B adds a relationship where User B follows User A. Merging these states would create a cycle, violating the constraint. Coordination prevents such circular references.\\

\textbf{3 Application Data Constraints (ADC):} This section provides examples of data constraints defined by applications that go beyond what can be enforced by built-in database constraints \cite{nodine1992cooperative}.

\textbf{Sequential Order}: Ensure data follow a specific sequence.  \textit{Non-Invariant Satisfaction Example:} In a task management application, Transaction A updates the \texttt{start\_date} of a task, and Transaction B updates the \texttt{end\_date}. If these transactions occur simultaneously and the \texttt{start\_date} ends after the \texttt{end\_date} when merged, the sequence is violated. Coordination ensures the proper order of operations.

\textbf{Conditional Value Constraints}: Values in certain fields depend on others in the data structure. \textit{Non-Invariant Satisfaction Example:} In a customer management system, a VIP customer must have an account balance above \$10,000. Transaction A modifies the customer's VIP status, and Transaction B deducts from their balance. Without coordination, the merged result could reflect a VIP status with an insufficient balance, violating the constraint.

\textbf{Session Data Consistency}: Ensures that data remains stable and isolated within a user’s session, preventing modifications by other users during that session. \textit{Non-Invariant Satisfaction Example:} In an online shopping cart, if Transaction A holds the session while adding an item and Transaction B from another user modifies the cart’s contents simultaneously, merging these actions could result in an inconsistent cart state. Coordination helps maintain session data stability and integrity.\\

\textbf{4 Application Process Constraints (APC)}: This section provides examples of constraints for application procedures or workflows, focusing on the sequence and conditions of operations rather than the data state itself \cite{biliris1994asset, nodine1992cooperative}.

\textbf{Sequence Requirements}: Define that specific actions must occur in a specific order. \textit{Non-Invariant Satisfaction Example:} In an order processing system, payment must be confirmed before delivery is initiated. Transaction A processes the delivery, while Transaction B confirms payment from the same order. If merged without coordination, delivery could occur before payment, violating the process requirements.

\textbf{Grouped Actions}: Enforce that certain operations must happen together. When adding a new student, they must be inserted into both the department and dorm tables. \textit{Non-Invariant Satisfaction Example:} If Transaction A inserts the student into the department table and Transaction B inserts them into the dorm table, merging without coordination may leave an inconsistent state where the student is only present in one table.

\textbf{Branching/Conditional Logic Constraints}: Define constraints based on conditional paths in the data or process flow.  \textit{Non-Invariant Satisfaction Example:} If Transaction A updates \( x \), and Transaction B adjusts \( y \) based on the new value of \( x \) (e.g., if \( x > 0 \), \( y \times 2 \); otherwise, \( y + 2 \)). The existence of x decides the value of y, and they cannot be run concurrently since Transaction B needs to wait for the result of Transaction A.

\section{I-Satisfaction Check, Completeness of Query, Invariants}

\textbf{Scenario 1}: Assume that transaction type A is deducting \( x_1 \) USD from account balance of row \( y_1 \), and transaction type B is deducting \( x_2 \) USD from account balance of row \( y_2 \). \textbf{Invariant}: account balance \( > 0 \).

In a web development environment with a relational database, queries within a transaction are represented as SQL query templates, which may require users to input specific parameters. This means that transactions can have different levels of completeness depending on whether all required parameters are specified. For example, in Scenario 1, parameters like \(x_1\), \(x_2\), \(y_1\), and \(y_2\) may or may not need to be defined by the user, depending on application logic. In more complex transactions, values for \(x_1\), \(x_2\), \(y_1\), and \(y_2\) might be dynamically retrieved from a \texttt{SELECT} query.

On the other hand, we cannot assume that every developer or user has prior knowledge of all constraints. We will explore how varying levels of completeness in query information affect dependency checks for ensuring consistency. For Scenario 1, we explained the granularity of dependency checks, and in Table 2, we summarized dependency checks for other constraints based on actions and completeness of information.

\textbf{Complete Query (No User Input), with Invariant Information:} When no user input is required, the exact parameters for each transaction are known in advance. For example, transaction type A always deducts 2 USD from account row 1, and transaction type B also deducts 2 USD from account row 1. In this case, without knowing the balance, these two transactions require coordination. We can simply hard-code transaction type B to wait for type A to finish whenever they interact. Here, the second transaction depends on the first only if they both deduct from\textbf{ the same field in the same row}, which has an invariant that ensures the value remains positive.

\textbf{Complete Query (With User Input), with Invariant Information:} When user input is required, dependencies between templates cannot be hard-coded. Instead, we can only identify the row affected once the user submits the necessary parameters. For example, if transaction A deducts 2 USD from account row \( y_1 \) and transaction B deducts 2 USD from account row \( y_2 \), we need to dynamically check if \( y_1 \) and \( y_2 \) refer to the same row based on user inputs. If they do, coordination is required.

\textbf{Partial Query with Invariants Info}: If the row number from transaction 1 is read from the system, then transaction 1 is an incomplete partial query until the read is finished. To check the invariant between two transactions requires “locking” the entire column. In this case, if they deduct a field in the same column with an invariant requiring a positive value, the second transaction depends on the first transaction.

If the row number for transaction 1 is retrieved from the database system, then transaction 1 is considered an incomplete or partial query until this read operation is completed. To check the invariance between two transactions, it may be necessary to “lock” the entire column. In this case, any new deduction transaction depends on the first deduction transaction if they both deduct a field in\textbf{ the same column}, with an invariant requiring that the value remain positive.

\textbf{Query with No Invariants Info}: In Scenario 1, if we do not know where the invariant has been applied to the table, the only solution is to "lock" the table. If the second transaction modifies the same table, it should wait and be held. In cases with foreign key constraints, where there is no referencing/referenced table information, strict serializability should be enforced.

In Scenario 1, if we do not know where the invariant has been applied to the table, the only solution is to “lock” \textbf{the entire table}. If the second transaction modifies the same table, it must wait and be held. In cases involving foreign key constraints, where no information about referencing or referenced tables is available, strict serializability should be enforced.

\textbf{Row/Column/Table/DB?}: Is there a universal rule for determining whether a dependency check depends on operations to the same row, column, table, or entire database? The answer is no; it depends on the specific operations and invariant pairs involved. Strictly speaking, an Invariant Satisfaction Check (IC) acts as a form of locking, but it leverages knowledge of invariants to narrow the scope of what needs to be locked. When query and constraint information is incomplete, a broader scope of “locking” is required.

\textbf{Consistency Validation:} It is important to note that, in invariant confluence, all transactions have been committed within a branch. All intermediate and final states are valid under database constraints. In our setting, however, transactions are buffered and thus not immediately committed for constraints validation. The logical dependency check only identifies dependencies between transactions; it does not determine whether a transaction can ultimately commit while preserving consistency. 

A buffered suspicious transaction only holds dependent transactions to prevent them from reducing the buffered suspicious transaction's likelihood of committing successfully. For instance, in Scenario 1, a second transaction might consume a positive balance, which could make the first (deducting) transaction inconsistent. However, whether the first transaction can be committed while maintaining system consistency depends on the current balance. The consistency of an individual transaction is ultimately verified and enforced by the database's validation logic (not needed for application-level constraints).

In the case of the dependent-upon (buffered) transaction is found inconsistent after physical constraints validation, it will be discarded, and the dependent transaction will not be affected, since the dependent-upon transaction does not change any database state.

\textbf{Physical Dependency Check \& Logical Dependency-Check:} We knew that two concurrent decrement transactions may lead to a database state that violates database consistency. Rather than committing transactions to the database and relying on built-in constraints validation, we use logical reasoning to identify dependencies. There are several reasons we prefer a Logical Check over a physical check. 1)  A logical check allows us to extend and support dependency checks that are not natively built into the system. 2) Dependency checking through physical transaction simulation is less efficient and will be explained in the discussion section.

    \begin{table*}[t!]
        \centering
        \renewcommand{\arraystretch}{1} % Adjust row height for better readability
        \resizebox{\textwidth}{!}{
\begin{tabular}{|>{\raggedright\arraybackslash}p{0.18\linewidth}|>{\raggedright\arraybackslash}p{0.12\linewidth}|>{\raggedright\arraybackslash}p{0.12\linewidth}|>{\raggedright\arraybackslash}p{0.18\linewidth}|>{\raggedright\arraybackslash}p{0.18\linewidth}|>{\raggedright\arraybackslash}p{0.18\linewidth}|}
\hline
\textbf{Invariants} & \textbf{Transaction 1 Action} & \textbf{Transaction 2 Action} & \textbf{Complete Query with Invariant} & \textbf{Partial Query with Invariant: $T_1$ missing row/column information} & \textbf{Without Invariant: do not know which row/columns has constraint} \\ \hline

\multicolumn{6}{|l|}{\textbf{Database Constraints}} \\ \hline
Uniqueness & Update/Insert & Update/Insert & Update/Insert the same column & $T_2$ Update/Insert the same table as $T_1$ & $T_2$ Update/Insert the same table as $T_1$ \\ \hline
Foreign Key & Delete key in a referenced table with no cascading delete & Insert to the referencing table & Insert row referencing the deleted row of the referenced key & Insert row referencing any key in the referenced table & Insert row referencing any key in the referenced table \\ \hline
Auto-Increment & Insert & Insert & Insert the same table & Insert the same table & Insert the same table \\ \hline
Check (less than) & Increment & Increment & Increment the same field & Increment the same column & Increment the same table \\ \hline
Check (more than) & Decrement & Decrement & Decrement the same field & Increment the same column & Increment the same table \\ \hline

\multicolumn{6}{|l|}{\textbf{Abstract Data Types Constraints}} \\ \hline
Counter (less than) & Increment & Increment & Increment the same field & Increment the same column & Increment the same table \\ \hline
Counter (more than) & Decrement & Decrement & Decrement the same field & Increment the same column & Increment the same table \\ \hline
Set, List, Map: size = & Mutation to the ADT & Mutation to the ADT & Mutate the same table (one table is a set) & Mutate the same table (one table is a set) & Mutate the same table (one table is a set) \\ \hline
Tree: Child has a Parent & Delete the parent & Insert a child & (A row is a node) Insert row as the child of the deleted row & Inserted row is in the same table as the deleted row & Inserted row is in the same table as the deleted row \\ \hline
Graph: No Circular Reference & Insert a reference $b$ & Insert $b$ referencing $a$ & (A row is an edge) Insert in the same reference column & (A row is an edge) Insert in the same reference column & (A row is an edge) Insert in the same table \\ \hline

\multicolumn{6}{|l|}{\textbf{Application Logic Data Constraints}} \\ \hline
Sequential Order: In a row, the start date is smaller than the end date & Insert/Update & Insert/Update & Insert/Update the same row & Insert/Update the same table & Insert/Update the same table \\ \hline
Conditional Value Constraints: If a customer’s status is VIP, their account balance must be above a certain threshold & Insert/Update (VIP status) & Insert/Update (deduct balance) & Inserted balance row constrained by the VIP row & Inserted balance row constrained by any row in the VIP status table & Inserted balance row constrained by any row in the VIP status table \\ \hline
Session Data Consistency: Data states do not change unexpectedly within a single session & Action from user 1 & Action from user 2 & Different users work in the same table & Different users work in the same table & Different users work in the same table \\ \hline

\multicolumn{6}{|l|}{\textbf{Application Logic Process Constraints}} \\ \hline
Sequence Requirements: Payments happen before deliveries & Insert (order is LLM generated) & Insert (delivery should wait) & Delivery insert row corresponds to a specific payment insert & Delivery insert corresponds to any insert in the payment table & Delivery insert corresponds to any insert in the payment table \\ \hline
Grouped Actions: Adding a student includes inserting them in both the department table and the dorm table & Insert & Insert & \multicolumn{3}{c|}{One must wait until the other appears} \\ \hline
Branching/Conditional Logic Constraints: Transaction $A$ updates $x$; transaction $B$ if $x > 0$, $y \times 2$, else $y + 2$ & Update (transaction 1) & Update (transaction 2) & \multicolumn{3}{c|}{Transaction $B$ needs to wait until transaction $A$ is finished} \\ \hline

\end{tabular}
}
\caption{Comparison of Transaction Actions under Different Query and Invariant Conditions}
        \label{tab:comparison-table}
    \end{table*}

\section{Transaction Manager}

In this section, we consider Scenario 2: transaction type A adds \( x_1 \) USD to the account balance of the row \( y_1 \), and transaction type B deducts \( x_2 \) USD from the account balance of the row \( y_2 \). Invariant:\textbf{ }The account balance must remain \( > 0 \).

\textbf{Isolation Coordination:} Isolation in this paper refers to the "I" in the ACID properties of a database system, ensuring that transactions are processed independently to avoid conflicts \cite{gray1981transaction}. Isolation coordination involves concurrency control to guarantee the serializability of transactions. When two transactions have write-write, read-write, or write-read conflicts and are running concurrently, isolation coordination ensures that the outcome is equivalent to executing these operations in a specific sequential order. In our framework, the database manages isolation, while the middleware we proposed is responsible solely for maintaining consistency.

\textbf{Consistency Coordination:} We have previously discussed consistency and dependency. To ensure system consistency while allowing for the potential removal of suspicious transactions, coordination between transactions is necessary. For instance, if we have two transactions that are both type A (increment operations) in a given scenario, there is no dependency, and no coordination is required. Similarly, if one transaction is an increment and the other a decrement, no coordination is needed. However, if both transactions are decrement operations, a dependency exists, and the second transaction should wait until the first transaction is complete.

Since we cannot assume a specific isolation level in the database system, we consider that transactions without coordination constraints may be committed concurrently.

\subsection{Processes vs. Transaction Manager}

We may need a machine to coordinate transactions. In theory, all uncommitted transactions could be managed in separate processes and threads, which would host transactions waiting for locks, perform simulations, or carry out logical dependency checks between new transactions and all buffered suspicious or compensating transactions. However, this approach has high bandwidth costs. More importantly, it is impractical in a long-lived transaction scenario, as we discussed, where processes might need to wait for days for user reviews. A more efficient solution is to build a transaction manager that can track transactions, manage their state, and commit them as appropriate.

\subsection{Transaction Manager Environment}

\begin{center}
\begin{tabular}{ccc}
Committed & Buffered & New \\ \hline
T1, T2, T3 & (T4) (T5) (T6) & T7
\end{tabular}
\end{center}

In the above diagram, we provide an example of dependency checking, and there are three types of transactions:

\textbf{Committed transactions} constitute the current state of the database. \textbf{Buffered transactions} include either suspicious transactions or compensating transactions to support transaction separation/removal. Buffered transactions also include those that depend on other previous buffered transactions. \textbf{New transactions} are added to the buffer if they are marked as suspicious. The system then checks whether a new transaction depends on any buffered transaction. If there are \( N \) buffered transactions, each new transaction requires \( N \) dependency checks. When a dependency is identified, it is recorded accordingly, as characterized in Table 1.

\subsubsection{How to manage/buffer transactions:}

The transaction manager handles all incoming new transactions serially, one at a time. Each transaction is evaluated against Invariant-Satisfaction for dependency check. If there is no dependency: The transaction is materialized (committed) into the database. If dependencies exist: The transaction is retained in the buffer for future processing.

\subsubsection{How to check dependency:}

Assume we have two transactions, T1 and T2. Go to Table 1 and check whether these two have actions paired in the same row of the table. Then, based on the completeness of the information at the time of commit, they can get to know the dependency criteria. For example, T1 is UPDATE key = 1 from table1 where id = 1, AND T2 is Update key = 2 from table1 where id = 1, and we know the "key" field has a uniqueness constraint. Then we go to table 1, go to row 1, and the "Complete Query with Invariant" column since we see it mentioned the same actions as T1 and T2, and we have complete query information that we know those transactions are working on the same "key" field which has uniqueness constraints. Then, dependency and coordination are needed. If the transaction has missing parameters related to columns or rows because it "read" real-time information from the database, we should check "Partial Query with Invariant: missing row/column information." If we do not know which rows/columns/tables have this constraint, but we know there may be one somewhere, we should go to "Without Invariant: do not know which row/columns has a constraint."

\subsubsection{How to represent the interdependence between transactions:}

Transactions are retained in the buffer not just because they are suspicious but because their dependencies are logged in the Dependency Matrix. The Dependency Matrix tracks relationships between transactions, indicating interdependence, as shown in the table below.

\begin{figure}
    \centering
    \includegraphics[width=1\linewidth]{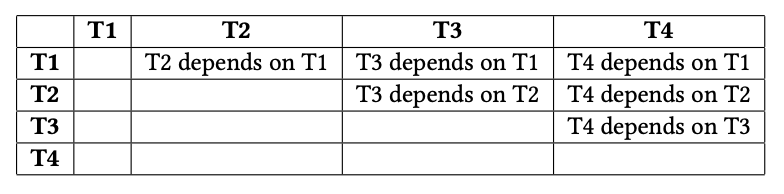}
    \caption{Dependency Matrix}
    \label{fig:Dependency-Matrix}
\end{figure}

\subsubsection{Algorithm for Adding a New Transaction}
This algorithm is implemented as \texttt{process\_transactions()}.

\noindent
\textbf{Input:} DB, Transaction Buffer, Dependency Matrix DM, New transaction T \newline
\textbf{Output:} Updated DB, Updated Transaction Buffer, Updated Dependency Matrix \newline

If T is suspicious, it will be added to the DM and wait for administration processing. Then, the transaction manager checks T's dependency against all existing transactions in the dependency matrix. For any transaction with more than one dependency identified, it specifies its dependency in the dependency matrix.

If T is not suspicious, Then the transaction manager checks T's dependency against all existing transactions in the dependency matrix. Only after the dependency is identified will it be added to the transaction buffer, and its dependency will be specified. For non-suspicious transactions, if no dependency is detected, they will be committed to the DBMS.

\subsubsection{Algorithm for Materializing a Buffered Transaction}

This algorithm is implemented as \texttt{check\_for\_materialization()}.

\noindent
\textbf{Input:} DB, Transaction Buffer, Dependency Matrix DM \newline
\textbf{Output:} Updated DB, Updated Transaction Buffer, Updated Dependency Matrix \newline

\noindent
The Transaction Manager regularly checks the Dependency Matrix. For each transaction in the Dependency Matrix:
\begin{enumerate}
    \item First, check whether it is approved by the administrator or if it is not a suspicious transaction.
    \item Then, check the dependency table vertically to see whether the transaction is still waiting for others.
\end{enumerate}

\noindent
If both conditions (1 and 2) are satisfied, the Transaction Manager will:
\begin{itemize}
    \item Remove its horizontal dependencies from the Dependency Matrix.
    \item Release any transactions that depend on it.
    \item Commit the transaction to the DBMS.
\end{itemize}

\section{Experiment}

In our system setting, we characterize the availability of the middleware and web service pipeline by the buffered rate. While end-to-end performance—from the user’s request to transaction completion—may serve as a more comprehensive availability metric, it is influenced by the specific implementations of the transaction manager and middleware, which are not the primary focus of this paper. Also, database performance can impact end-to-end performance. Thus we assume that the database completes all transactions immediately once they exit the middleware. We only simulate transaction management and do not commit it to the database.
\[
\text{Buffered rate} = \frac{\text{Number of buffered transactions}}{\text{Number of total test transactions}}
\]

TPC-C (Transaction Processing Performance Council Benchmark C) \cite{tpcc} is a standardized benchmark for evaluating the performance of transaction processing systems. It simulates an order-entry environment, which includes operations such as processing new orders, payment transactions, order status checks, deliveries, and stock level monitoring. TPC-C is widely used to measure and compare the efficiency and scalability of database and transaction management systems under a complex, real-world workload.

We have a naive intuition that with more dependency between transactions, the buffered rate will increase. In a web service system, if the proportion of inter-dependent transactions is higher, the buffer rate will accordingly increase. Thus, a fixed distribution of transactions could give us a better experiment parameter control to explore other parameters that affect the buffered rate. This is also the reason we use TPC-C as the test framework because it assumes a roughly fixed distribution between 5 types of transactions. 

A natural intuition suggests that as dependencies between transactions increase, the buffered rate will also rise. In a web service system, a higher proportion of interdependent transactions leads to a corresponding increase in the buffered rate. Therefore, using a fixed distribution of transactions provides better control over experimental parameters, allowing us to explore other factors affecting the buffered rate. This rationale also underlies our choice of the TPC-C framework for testing, as it maintains a roughly fixed distribution across five transaction types.

We implement our test frame by adapting and rewriting a TPC-C benchmark tool (py-tpcc) \cite{py-tpcc}. It is originally used to generate TPC-C transactions, dock transactions to a database, and measure the throughput performance of the target database. We rewrite this package, dock the transactions to our transaction manager, and control the flow of transaction generation. In our experiments, we mainly control 2 factors, the interval between two LLM-generated suspicious transactions (SI), and the interval between two user reviews to accept or remove the suspicious transaction, (RI). For simplicity, we assume equal intervals for both RI and SI.\\

\begin{figure}[ht!]
    \centering
      % Third row of figures
    \begin{minipage}{0.48\linewidth}
        \centering
        \includegraphics[width=\linewidth]{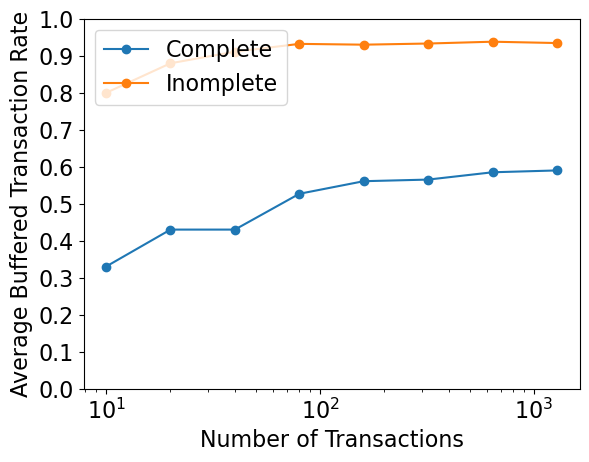}
        \caption{Conclusion 1.1: Without user reviews to reduce buffered transactions (RI = 50, SI = 5, and 20 trials per transaction length), \textit{the average buffered transaction rate is higher in scenarios with complete information compared to those without complete information.}}
        \label{fig:e5}
    \end{minipage}
    \hfill
    \begin{minipage}{0.48\linewidth}
        \centering
        \includegraphics[width=\linewidth]{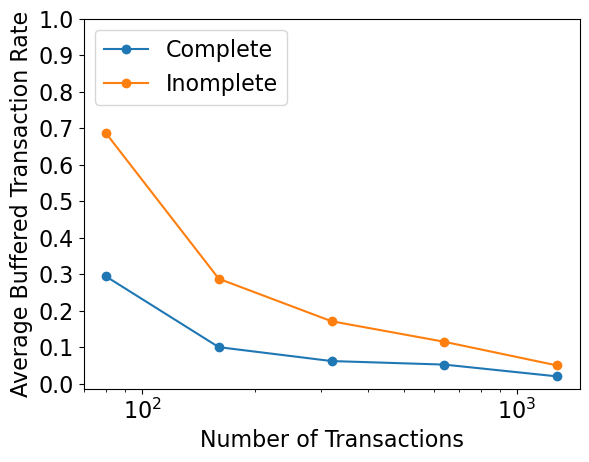}
        \caption{Conclusion 1.2: With user reviews at an 80\% rate (RI = 50, SI = 5, and 20 trials per transaction length), \textit{the average buffered transaction rate is higher in scenarios with complete information compared to those without complete information.}}
        \label{fig:e6}
    \end{minipage}
    % First row of figures
    \begin{minipage}{0.48\linewidth}
        \centering
        \includegraphics[width=\linewidth]{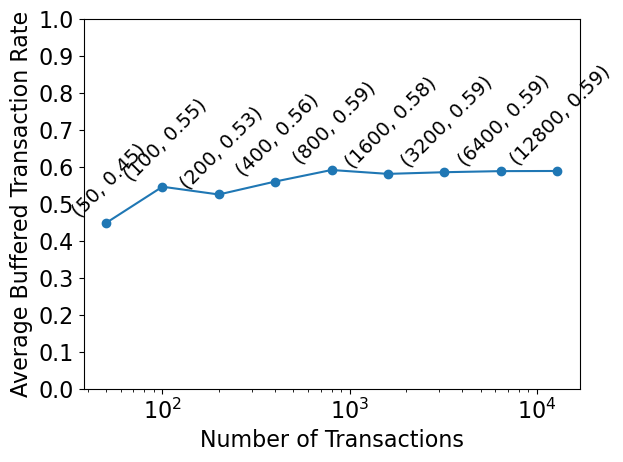}
        \caption{Conclusion 3.1: If the RI is infinite, the buffered rate does not change significantly as the number of transactions increases.}
        \label{fig:e1}
    \end{minipage}
    \hfill
    \begin{minipage}{0.48\linewidth}
        \centering
        \includegraphics[width=\linewidth]{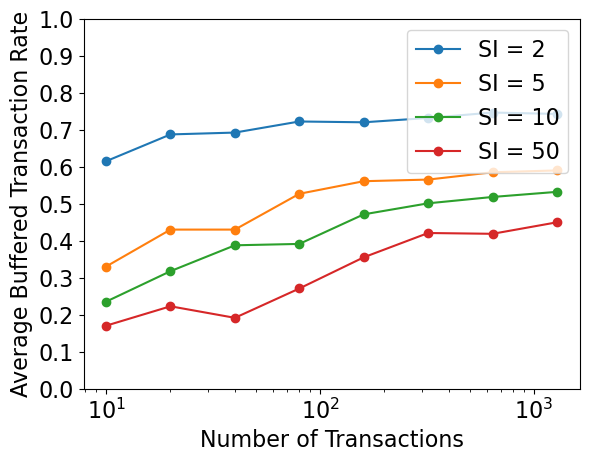}
        \caption{Conclusion 3.2: If the RI is infinite, a longer SI leads to a decrease in the average buffered rate.}
        \label{fig:e2}
    \end{minipage}
    
    \vspace{0.5cm} % Adjust space between rows

    % Second row of figures
    \begin{minipage}{0.48\linewidth}
        \centering
        \includegraphics[width=\linewidth]{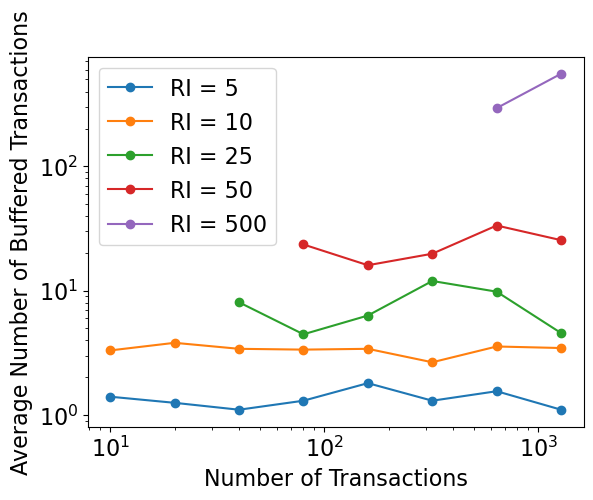}
          \caption{Conclusion 3.4: A shorter RI leads to a lower \textit{average number} of buffered transactions (SI = 5). Conclusion 3.5: When RI is fixed, the \textit{number of buffered transactions} remains relatively stable, even as the total number of transactions increases (SI = 5).}
        \label{fig:e3}
    \end{minipage}
    \hfill
    \begin{minipage}{0.48\linewidth}
        \centering
        \includegraphics[width=\linewidth]{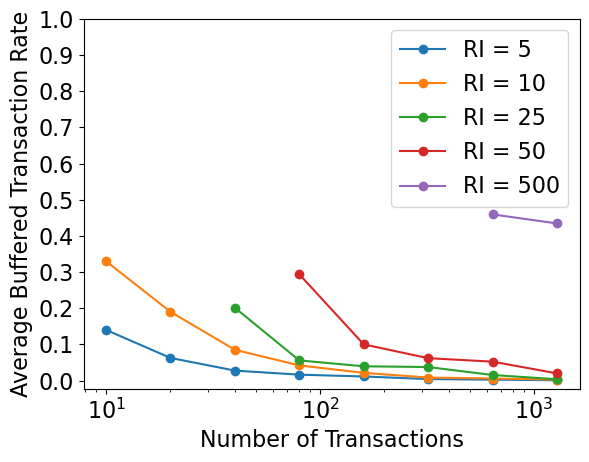}
        \caption{Conclusion 3.6: The buffered transaction rate decreases as the total number of transactions increases. (SI = 5)}
        \label{fig:e4}
    \end{minipage}
\end{figure}

\textbf{Question 1: Benchmark Comparison: how does the completeness of the query and invariant affect the buffered rate}

From our previous discussion, we understand that with higher information completeness to queries and invariants, we can assess transaction dependency with finer granularity. For example, in Scenario 1, with complete query and invariant information, we identify a dependency if the second transaction modifies the same field in the same row as the first transaction, assuming a "> 0" constraint on that field. However, without prior knowledge of invariants, we may need to block the entire system to ensure consistency for the first transaction.

We assume that a system without using invariant information can coordinate transactions in the TPC-C benchmark only through table-level locking. This is our baseline.

Therefore, we tested how dependency check granularity, under different levels of query and invariant completeness, affects the buffered rate. Our previous experiments used dependency checks based on Invariant Satisfaction. In TPC-C’s design, all invariant-related transaction queries are complete at the time of coordination checks, meaning a new transaction only depends on a buffered transaction if they modify the same field. To compare, we implemented a transaction manager without prior invariant knowledge using table-level dependency checks. This means that a second transaction is held or buffered as long as it modifies the same table as the first transaction.

\textbf{Conclusion 1.1}: When RI is infinite, and SI is set to 5, there is no user review to reduce buffered transactions. With 20 trials, \textit{the average buffered rate with field-level granularity dependency checks (complete invariants and query information) is lower than with table-level granularity checks (incomplete invariants and query information).} This suggests that with complete information about invariants and queries, a system can achieve lower buffered rates through finer-grained coordination compared to cases with no invariant information (see Figure \ref{fig:e5}).

\textbf{Conclusion 1.2}: If user reviews occur at a rate of 80\%, which removes 80\% buffered transactions, (RI = 50, SI = 5, with 20 trials for each transaction length), \textit{the average buffered transaction rate in the case with complete information is higher than in the case without complete information } (see Figure \ref{fig:e6}).

\textbf{Benchmark Conclusion 1.3}: Regardless of user reviews, the buffered rate in the baseline benchmark (without invariant information) is double that of the benchmark utilizing complete query and invariant information.
\\

\textbf{Question 2: Which type of transaction is most likely to cause congestion or hold new transactions, leading to an increase in the buffered rate?}

From the manual analysis and invariant confluence, in TPC-C, the dependency exists between New-order and New-order Transactions; and between Delivery and Delivery Transactions.\\

\textbf{Question 3: What factors related to suspicious transactions affect the transaction acceptance rate?}

\textbf{Conclusion 3.1} If the RI is set to infinity—meaning there is no review to remove buffered transactions that may congest new transactions—the buffered rate does not change significantly as the number of transactions increases.

In this scenario, we assume an SI of 5, where one out of every five transactions is labeled as an LLM-generated or suspicious transaction (e.g., True, False, False, False, False, True, …). We calculate all buffered rates 5 times and average the results. As shown in Figure \ref{fig:e1}, the average buffered rate remains steady at around 0.6 if the number of transactions is larger than 400, regardless of the increasing number of transactions.

\textbf{Conclusion 3.2} If the RI is set to infinity, a longer SI leads to a decrease in the average buffered rate.

We varied the interval between two LLM-generated transactions, testing SIs of 2, 5, 10, and 50 (e.g., one LLM-generated transaction for every 2 transactions, with 20 trials per SI). As shown in Figure \ref{fig:e2}, the test case with higher SIs results in a lower average buffered rate.

\textbf{Fact 3.3}: If the user reviews and either approves or rejects all transactions at any point, all buffered transactions will be processed, resulting in a buffered transaction rate of 0.

\textbf{Conclusion 3.4} A shorter RI results in a lower \textit{number} of average buffered transactions.
 
Since reviewing all transactions would result in a buffered rate of 0, we tested the effect of partially reviewing transactions by randomly removing 80\% of the transactions in the buffer. Assuming that human reviews occur less frequently than LLM-generated transactions, we set the RI (user review interval) to be a multiple of the SI (suspicious transaction interval), with values of (1, 2, 5, 10, 100) times the SI. For this test, we fixed the SI at 5, giving RI values of (5, 10, 25, 50, 500),  with 20 trials per RI.

We focused on cases where the number of transactions exceeded the RI to ensure human processing within each test case, omitting scenarios without human processing. Consequently, a few lines in Figure \ref{fig:e3} are truncated. Also, as shown in Figure \ref{fig:e3}, a shorter RI results in fewer buffered transactions.

\textbf{Conclusion 3.5}: When RI is fixed, the number of buffered transactions remains relatively stable, even as the total number of transactions increases.

\textbf{Conclusion 3.6}: Because the number of buffered transactions remains stable, the buffered transaction rate decreases as the total number of transactions increases (see Figure \ref{fig:e4}).\\

% \textbf{Question 4: How stable is GPT in generating transaction parameters?}

% (Only called the API once to generate 100 transactions.)

% \textbf{4.1 Correctness}:
% We checked the correctness of GPT-generated transaction parameters for:
% \begin{enumerate}
%     \item Inclusion of required values for each type of transaction.
%     \item Correct format of the parameters.
%     \item Generated parameters being within the required value range.
% \end{enumerate}

% \textbf{4.2 Distribution Comparison}:

% \textbf{Transaction Type Comparisons}:
% \begin{itemize}
%     \item \textbf{STOCK\_LEVEL}: Expected: 4\%, Actual: 0.00\%
%     \item \textbf{DELIVERY}: Expected: 4\%, Actual: 0.00\%
%     \item \textbf{ORDER\_STATUS}: Expected: 4\%, Actual: 10.00\%
%     \item \textbf{PAYMENT}: Expected: 43\%, Actual: 30.00\%
%     \item \textbf{NEW\_ORDER}: Expected: 45\%, Actual: 60.00\%
% \end{itemize}

\section{Discussion}

\subsection{Mechanism Comparison}
We have discussed several mechanisms and approaches for addressing our system setting. However, we also mentioned they are not a good choice for our questions. How do these mechanisms compare, and why do we prefer buffering suspicious or compensating transactions with the invariant-satisfaction dependency check?

\subsubsection{Naive Buffering}
In naive buffering, incoming removable transactions are stored in a buffer and committed or removed if they are reviewed by the admin users. The naive buffering strategy does not assume any prior knowledge about invariants. Essentially, suspicious transactions are moved to the end of the transaction history, with no coordination between new transactions and buffered transactions. Thus, there is a risk that a transaction deemed consistency-valid may fail to commit (Scenario 1), and naive buffering does not guarantee the commit of a buffered transaction.

\subsubsection{Transaction Simulation}
Transaction simulation periodically checks if a compensating transaction can be used to remove an erroneous transaction. When a new transaction arrives, we set up a sandbox environment and commit various combinations of the new and compensating transactions to a database snapshot. As long as all combinations preserve system consistency, the new transaction can be committed. This simulation approach ensures that any committed transaction can be removed if necessary using its corresponding compensating transaction.

If there are buffered compensating transactions and new transactions ready for undo simulation, let $N$ be the number of buffered transactions and $n$ the number of new transactions. $(2^N) \times n$ simulations may be needed to test all combinations. Meanwhile, our proposed buffering compensating/suspicious approach only requires $N \times n$ I-satisfaction dependency checks. Additionally, it’s important to note that simulation involving physical I/O is significantly more time-intensive than in-memory logical I-satisfaction checks.

\subsubsection{Buffering Suspicious Transaction Is Better than Buffering Compensating Transaction ?}
One key drawback of buffering compensating transactions is if a transaction is reviewed and removed by the user, the system actually interacts with the database twice for committing. It commits the original suspicious transaction and the undo compensating transaction. This doubles the transaction overhead if users remove all LLM-generated transactions. For buffering suspicious transaction approaches, no additional actions are required other than removing transactions from the buffer, and no database execution is needed.

\subsubsection{Buffering Compensating Transaction is better than Buffering Suspicious Transaction:}
Invariant satisfaction only provides guidance on whether two transactions need coordination if they are requested by users concurrently. However, invariant satisfaction does not provide any guarantee whether this transaction can be committed and satisfy database consistency.
We summarize the constraints validation of the above-mentioned approaches below:
\begin{itemize}
    \item \textbf{Pure Buffering:} Constraint validation is done at the point of committing or through extra application logic.
    \item \textbf{Simulation:} Constraint validation is tested at the simulation stage.
    \item \textbf{Buffering Suspicious Transactions with Invariant Satisfaction:} If transaction 1 depends on transaction 2, the transaction will be constraint-validated until transaction 1 has been reviewed by the user and submitted to the database. If transaction 2 is not suspicious, it needs to wait until transaction 1 is finished. Thus, constraint validation has been delayed significantly.
    \item \textbf{Buffering Compensating Transactions with Invariant Satisfaction:} The suspicious transaction can be committed and validated when the user requests. The corresponding compensating transaction is the same. There is no delay for the consistency validation check. 
\end{itemize}

In scenarios with many consistency-invalid transactions that should be aborted, buffering compensating transactions enables immediate abortion and removal of the suspicious transaction at request time: When all transactions are consistency-invalid, buffering suspicious transactions wait a long for user review and unnecessarily pile up the buffer with invalid transactions.

\subsection{Lock}
A dependency check based on invariant satisfaction is similar to locking in a database system. The dependency check identifies conflicts between two transactions and holds the second transaction until the first one is completed. In databases, locks are primarily used to ensure isolation between transactions. If there is a write-write (WW), write-read (WR), or read-write (RW) conflict between two transactions, the lock is applied to the necessary data, requiring one transaction to wait until the other releases the lock \cite{eswaran1976notions}. However, locks for isolation typically have fixed granularity, such as row-level, table-level, or system-level locks, which may be insufficient for maintaining consistency in complex coordination scenarios which has varying granularity requirements.

For example, In the context of ensuring consistency, 2PL may require a table-level lock. To be more specific, for consistency coordination required invariants like uniqueness, the former transaction must lock the whole table to make sure all later transactions will not insert any row to lower the former transaction’s consistency priority (shrinking the former transaction’s range of unique names) \cite{eswaran1976notions}. Holding a table lock is also true for a foreign key since if I want to delete a referenced foreign key tuple, I need to lock the whole referencing table to make sure no new transaction will reference the deleted referenced key. For Invariant >, <, the system may only need to lock the corresponding rows to make sure no other transaction can modify it and negatively affect the former transaction’s consistency priority. Since the lock-based approach does not have information to invariant beforehand, it is necessary to assume a table-based lock to guarantee consistency.

For example, a two-phase locking (2PL) might enforce a table-level lock. It is a good granularity choice for invariants like uniqueness, where the first transaction must lock the entire table to prevent subsequent transactions from inserting rows that could violate the uniqueness constraint. However, it is not necessary for CHECK invariants with conditions (e.g., \verb|>|, \verb|<|); the system may only need to lock specific rows to prevent other transactions from altering them and compromising consistency. This fixation on granularity leads to coordination overhead. 

% Similar to why we use transaction manager instead of processes for managing coordination between transactions: Since the suspicious transactions are long-lived and may wait more than hours to hear back from the admin, the lock may consume a huge amount of computing process resources and reduce the availability of the system. One advantage of this approach compared with pure buffering is that the lock can guarantee a consistent commit even after a long time as long as the transaction is consistency-valid at the time of being pushed to the database.

% \subsection{Strict Serialization}
% (Except for consistency priority, there is another consideration: strict serialization. If different orders of transactions lead to different final states, previous transactions have a higher priority in strict serialization, which means their modification will be first applied. This can be done by checking commutativity. This is not the main focus of our current topics.) 

% (Commutativity check is an approach to guarantee strict serializability. We assume that we will buffer all suspicious transactions. For new transactions, we will check whether different commit orders of buffered suspicious transactions can lead to the same state. Similar to invariant confluence check.)

\subsection{Strict Serialization and Commutativity in Transactions}
Transaction orders can lead to different database states if two transactions are not non-commutative \cite{eswaran1976notions}. If we assume there is a "correct" order of transaction, any change to this order may lead to an "incorrect" state. Our framework can also ensure a favorable and "correct" outcome by coordinating these transactions based on commutative checks \cite{ranjan2021version}.

For buffered suspicious transactions and new transactions, to ensure the final state is identical to that of the "correct" order, we can use a commutativity check to identify necessary coordination. For a new transaction, if it is not commutative with any of the buffered transactions, then dependencies exist between the new transaction and the non-commutative buffered transactions. The coordination mechanisms remain the same as previously discussed. This approach not only guarantees a consistent database state but also ensures that the final state is identical to the "correct" transaction order.

\subsection{Undo with Compensating Transaction}

A compensating transaction may not exist if there is no logical inverse for a given operation. For example, setting a value to zero eliminates any knowledge of its previous state, making it impossible to reverse mathematically. Similarly, sending a notification or email represents an external action that cannot be "unsent" or undone once it leaves the system. This absence of an inverse action means that the system cannot revert these transactions in a straightforward manner, as the original state or condition cannot be restored.

Cascading compensating transactions occur when other transactions use the results of the current transaction \cite{garcia1987sagas}. If one transaction depends on or builds upon the outcome of another, reversing the initial transaction requires creating compensating transactions for all dependent ones. This cascade can become increasingly complex as each linked transaction must be undone or adjusted to maintain consistency.

\subsection{Related Works}
Undoing a transaction involves two key components: determining how to reverse the transaction’s effects and ensuring that compensating actions maintain system consistency. Traditional approaches (e.g., compensating transactions, ACTA, sagas) have largely addressed the first aspect—how to perform the compensation. However, the approach described here addresses both. Previous studies have explored ensuring consistency for long-lived transactions under weaker serializability assumptions.

Certain work has discussed verifying whether a compensating transaction can be applied based on write-read dependencies and introduced the notion of r-soundness, a weaker form of consistency ensuring that compensated or rolled-back states satisfy database constraints \cite{korth1990formal}. While that research established the concept, our method provides an exact procedure for verifying r-soundness through Invariant-C.

An earlier analysis addressed concurrency control and recovery together but focused on short-lived operations and traditional recovery mechanisms, not on long-lived transactions or compensating strategies \cite{schek1993towards}.

Additional studies employed abstract data types (ADTs) to manage long-lived transactions \cite{speegle1992quantifying}. This approach sometimes permitted concurrent execution without violating constraints, conceptually aligning with our integrity constraints (IC). However, while those works utilized alternative structures and focused less on preventing constraint violations, our approach employs a matrix-based mechanism explicitly designed to maintain database constraints.

Other efforts demonstrated that certain compatible transactions can commit without coordination if their semantics are aligned \cite{garcia1983using}, a concept similar to IC, yet they did not offer a comprehensive mechanism to identify and handle all potential conflicts as our IC-based solution does.

Further research categorized interactions among concurrent transactions using semantic considerations, discussing semantic and commutative compatibilities that maintain integrity and allow for free interleaving when operations result in equivalent states \cite{natrajan1999resolving}. However, these approaches did not focus on systematically enforcing database constraints.

Finally, some papers have summarized various serializability and correctness models—such as Predicate-wise Serializability (PSR), Quasiserializability (QSR), Cooperative Serializability (CoSR), and Setwise Serializability (SWSR). These models are group-based and restrict uncoordinated operations to occur only between groups rather than at the individual transaction level \cite{ramamritham1996taxonomy}. PSR, while conceptually related to IC, has been shown to improve the availability of long-lived transactions but only allows concurrent transactions between groups \cite{rastogi1993correctness}. This is stricter than r-soundness. In contrast, our Invariant-C approach allows for greater flexibility at the transaction level while still preserving global integrity.

% \subsection{challenges for fully dynamic invariant confluence check}

\begin{table*}[ht]
    \centering
    \caption{Comparison of Possible Transaction Management Strategies}
    \resizebox{\textwidth}{!}{  % Resize the table to half of the page width
        \begin{tabular}{|c|c|c|c|c|}
            \hline
            \textbf{Strategy} & \textbf{Transaction Required?} & \textbf{Invariant + Transaction Required?} & \textbf{Guaranteed Commit?} & \textbf{Physical Check?} \\
            \hline
            Naive Buffer & \checkmark & & & \checkmark \\
            \hline
            Simulation & \checkmark & & \checkmark & \checkmark \\
            \hline
            Buffer Suspicious Transactions & & \checkmark & \checkmark & \\
            \hline
            Buffer Compensating Transactions & & \checkmark & \checkmark & \\
            \hline
            Commutativity + Buffer & \checkmark & & \checkmark & maybe \\
            \hline
        \end{tabular}
    }
\end{table*}

\section{Conclusion}
In this work, we characterized the workflow of interactions between LLMs and databases in the presence of semantic errors. We introduced the concept of Invariant Satisfaction, which identifies the necessary coordination between long-lived, buffered transactions and new transactions. This ensures that new transactions do not lead to states where buffered transactions become irrecoverable. Invariant Satisfaction extends beyond database constraints to include application logic constraints, dynamically reducing coordination requirements based on the completeness of constraint and query information. We proposed a middleware framework for coordinating undo-able LLM-generated transactions that can be integrated into existing enterprise systems with minimal modifications. 

This paper provides a robust solution for managing undo-able long-lived transactions and guarantees consistency. For system researchers, this study introduces an interactive paradigm between LLMs and databases, featuring an "undoing" mechanism to handle incorrect operations while maintaining consistency. For system engineers, it offers a middleware design that efficiently incorporates undo-able LLM transactions into current workflows, ensuring reliability with minimal disruption.

% \begin{acks}
%  This work was supported by the [...] Research Fund of [...] (Number [...]). Additional funding was provided by [...] and [...]. We also thank [...] for contributing [...].
% \end{acks}

%\clearpage

\bibliographystyle{ACM-Reference-Format}
\bibliography{sample}

\end{document}